\documentstyle[12pt, axodraw]{article}
\newcommand{\AmS}{{\protect\the\textfont2
A\kern-.1667em\lower.5ex\hbox{M}\kern-.125emS}}

\def\be{\begin{equation}}
\def\ee{\end{equation}}
\def\bea{\begin{eqnarray}}
\def\eea{\end{eqnarray}}

\begin{document}
\begin{titlepage}

\begin{center}
{\bf The gluon Reggeization in perturbative  QCD at
NLO$^{~\ast}$}
\end{center}
\vskip 0.5cm \centerline{V.S.~Fadin$^{\dag}$} \vskip .3cm
\begin{center}
{\sl Institute for Nuclear Physics, 630090 Novosibirsk, Russia\\
and Novosibirsk State University, 630090 Novosibirsk,
Russia}
\end{center}
\vskip 1cm


\begin{abstract}
 Compatibility of the Reggeized form of QCD
multi-particle amplitudes with the  s-channel unitarity
requires fulfilment  of an infinite number of the
"bootstrap" relations.   On the other hand, it turns out
that fulfillment of all  these  relations ensures the
Reggeized form of energy dependent radiative corrections
order by order in  perturbation theory. It is extremely
nontrivial, that all these relations are fulfilled if the
Reggeon vertices and trajectory satisfy several bootstrap
conditions.  The full set of these conditions in the
next-to-leading order  was derived in the last year and
the ultimate condition was shown to be satisfied
recently. It means that the Reggeization hypothesis is
proved now in the next-to-leading approximation.

\end{abstract}
\vskip 7.5cm \vfill \hrule \vskip 0.3cm \noindent
$^{\ast}${\it Work supported in part  by the Russian Fund
of Basic Researches, grant 04-02-16685.} \vfill $
\begin{array}{ll} ^{\dag}\mbox{{\it e-mail address:}} &
\mbox{FADIN@INP.NSK.SU}\\
\end{array}
$
\end{titlepage}
\vfill
\eject

\section{Introduction}

One of remarkable properties of Quantum Chromodynamics is
{the gluon Reggeization}. Non vanishing in the high
energy limit cross sections are related to gluon
exchanges in cross channels. Therefore the gluon
Reggeization is extremely important for the description
of QCD processes at { high energy $\sqrt s$}. In
particular, this  phenomenon appeared as the basis of the
{BFKL approach} \cite{BFKL} to the description of high
energy processes. {It was proved} \cite{BLF} in the
{leading logarithmic approximation} (LLA), when only the
leading terms {$~(\alpha _S\ln s)^n$} are summed. Owing
to this the BFKL approach was grounded in the LLA. Now
the approach is intensively developed in the {next-to
leading approximation} (NLA), when the terms {$~\alpha
_S(\alpha _S\ln s)^n$} are also summed. {In this
approximation the gluon Reggeization  remained a
hypothesis till now}. Evidently, its proof is extremely
desirable. The proof is especially necessary because of
appearance of statements about existence of contributions
violating the Regge ansatz at three loop level
\cite{Kucs}. Now the desired proof is completed.

\section{The Reggeization hypothesis}

The hypothesis determines QCD amplitudes in the
{multi-Regge kinematics} -- MRK (at that the Regge
kinematics is considered as a particular case of the
MRK).  {MRK} is the kinematics  where all particles have
{limited (not growing with $s$) transverse momenta} and
are combined into jets with {limited invariant mass of
each jet and large} (growing with $s$) {invariant masses
of any pair of the jets}. At {leading order (LO)} only
gluons can be produced and each jet is actually a gluon.
 At {next-to-leading order (NLO)} a jet can contain a couple
of partons (two gluons or quark-antiquark pair). Such
kinematics is called also {quasi multi-Regge kinematics
(QMRK)}.

The MRK gives {dominant contributions to cross sections}
of QCD processes at high energy $\sqrt s$. In
perturbation theory these contributions are related to
exchanges of the gluon quantum numbers in cross channels
with fixed (not increasing with $s$) momentum transfers
$q_i$. The hypothesis is based on the calculations of QCD
amplitudes. Despite of a great number of contributing
Feynman    diagrams it turns out that at the Born level
{in the MRK amplitudes acquire a simple factorized form}.
Quite uncommonly that {radiative corrections} to these
amplitudes don't destroy this form, and their energy
dependence is given by {simple Regge factors}
${s_i}^{\omega(q_i)}$, where $s_i$  are  invariant masses
of couples of neighbouring jets and $\omega(q)$ can be
interpreted as  a shift of gluon spin from unity,
dependent from momentum transfer $q$. This phenomenon is
called {gluon Reggeization} and $\omega(t)$ is called
gluon Regge trajectory (although actually the trajectory
is $ j(t) = 1 + \omega(t))$. The Reggeization hypothesis
affirms that
\begin{equation}
\Re {\cal A}^{ A^{\prime}   B^{\prime} +n}_{AB}=2p^+_A
p^-_B\, \Gamma_{ A^{\prime}A} \left( \prod_{i=1}^n
 \frac{e^{\omega(q_i)(y_{i-1}-y_i)}}{q^2_i}\gamma^{J_i}(q_i,q_{i+1})
 \right) \frac{e^{\omega(q_{n+1})(y_{n}-y_{n+1})}}{q^2_{n+1}}\Gamma
_{ B^{\prime} B}. \label{A 2-2+n}
\end{equation}
Here $\Re$ means a real part; ${\cal A}^{A^{\prime}
 B^{\prime} +n}_{AB}$ is the amplitude for production of
jets $ A^{\prime},\;J_1,\;.....J_n,\; B^{\prime}$,
strongly ordered in rapidity space (see Fig.1);
$\Gamma_{P^{\prime} P}$ are the scattering vertices,
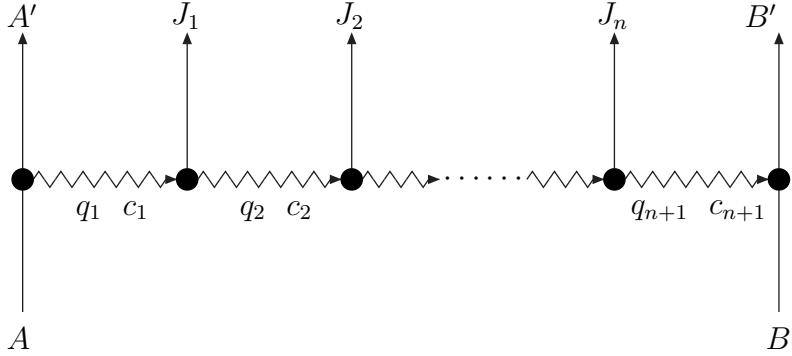
\begin{figure}[htb]
\begin{center}
\begin{picture}(200,120)(35,-80)

\LongArrow(-8,-2)(-8,48) \ZigZag(-4,-6)(46,-6){3}{6}
\ArrowLine(46,-6)(50,-6) \GCirc(-8,-6){4}{0}
\Line(-8,-56)(-8,-10) \Text(-14,56)[l]{$ A^{\prime}$}
\Text(-14,-66)[l]{$A$} \Text(12,-18)[l]{$q_1~~c_1$}

\GCirc(54,-6){4}{0} \LongArrow(54,-2)(54,48)
\ZigZag(58,-6)(108,-6){3}{6} \ArrowLine(108,-6)(112,-6)
\Text(48,56)[l]{$J_1$} \Text(74,-18)[l]{$q_2~~c_2$}

\GCirc(116,-6){4}{0} \LongArrow(116,-2)(116,48)
\ZigZag(120,-6)(145,-6){3}{3} \ArrowLine(145,-6)(149,-6)
\Text(151,-6)[l]{$\cdots \cdots$}
\ZigZag(182,-6)(207,-6){3}{3} \ArrowLine(207,-6)(211,-6)
\Text(110,56)[l]{$J_2$}

\GCirc(215,-6){4}{0} \LongArrow(215,-2)(215,48)
\ZigZag(219,-6)(269,-6){3}{6} \ArrowLine(269,-6)(273,-6)
\Text(209,56)[l]{$J_n$}
\Text(222,-18)[l]{$q_{n+1}~~c_{n+1}$}

\GCirc(277,-6){4}{0} \LongArrow(277,-2)(277,48)
\Line(277,-56)(277,-10) \Text(265,56)[l]{$ B^{\prime}$}
\Text(273,-66)[l]{$B$}

\end{picture}
\caption{Schematic representation of the process
$A+B\rightarrow A^{\prime}+J_1+\dots+B^{\prime}$ in the
MRK.}
\end{center}

\end{figure}
\noindent  i.e. the effective vertices for $P \rightarrow
P^{\prime}$ transitions due to interaction with Reggeized
gluons; $\; \gamma ^{J_{i}}(q_{i},q_{i+1})$ are the
production vertices, i.e. the effective vertices for
production of jets $J_i$ with momenta $k_i=q_i-q_{i+1}$
in collisions of Reggeons with momenta $q_{i}$ and
$-q_{i+1}$; $q_0=p_{A}-p_{ A^{\prime}}$, $q_{n+1}=p_{
B^{\prime}}-p_{B}$. We use light cone vectors $n_1$ and
$n_2$, $n_1^2=n_2^2=0; \;\; (n_1n_2)=2$ and denote
$p^\pm=(pn_{2,1})$. It is assumed that initial momenta
$p_A$ and $p_{B}$ have predominant components $p_A^+$ and
$p_B^-$.  For generality we do not assume that transverse
to the $(n_1, n_2)$ plane components
$p_{A\bot},\;\;p_{B\bot}$ are zero. Moreover,  $A$ and
$B$, as well as $A^{\prime}$ and $B^{\prime}$, can
represent jets. In (\ref{A 2-2+n}) $y_i$ are jet
rapidities,  $y_i=\frac{1}{2}\ln\left(k^+_i/k^-_i\right)$
for $i=1, ...n$, $y_0=
y_A\equiv\ln\left(p^+_A/q_{1\bot}\right) $,
$y_{n+1}=y_B\equiv
\ln\left(q_{(n+1)\bot}/p^-_{B}\right)$. We use positive
Euclidean metric for transverse components.

Note that Reggeons, as well as gluons, belong to colour
octet, so that the vertices carry colour indices. For
simplicity, we omit these indices when they are not
necessary for understanding.

The factorized form of QCD amplitudes in the {MRK} was
proved at the {Born level} using the $t$--channel
unitarity and analyticity.  Their {Regeization}  was
firstly derived in the LLA on the basis of the direct
calculations at  the {three-loop level for elastic
amplitudes} and the {one-loop level for one-gluon
production amplitudes}. Later {it was proved \cite{BLF}
in the LLA for all amplitudes at arbitrary number of
loops} with the help of {bootstrap relations}. {At NLO
the Reggeization remained a hypothesis till now.}

{The hypothesis is extremely powerful} since an infinite
number of amplitudes is expressed in terms of the gluon
Regge trajectory and several Reggeon vertices.

\section{{\bf  Idea of the proof of the hypothesis}}

A basic idea of the proof is based on use of  the
{$s$-channel unitarity}.   In order to realize the idea
we need {to express  real parts of amplitudes in terms of
their  $s$-channel discontinuities}. It is not difficult
to do for elastic amplitudes.  Unfortunately, it is quite
not so for {inelastic amplitudes}.

{But if in the MRK we confine ourselves to the NLO, the
situation changes drastically:} \cite{V.F.02} {the
discontinuities} (more precisely, real parts of definite
combinations of the discontinuities) {are related to the
derivatives of the amplitudes over jet rapidities} : \be
\sum_{l=j+1}^{n+1}\Delta_{{jl}}
-\sum_{l=0}^{j-1}\Delta_{{lj}} \, =\,\frac{\partial
}{\partial y_{j}}\Re \; \left[e^{y_B-y_A}{\cal
A}_{AB}^{A'B'+n} (y_{i})\right],\label{deriv} \ee where
\be \Delta_{{jl}} \,=\, e^{y_B-y_A}\;\Re\;\frac{1}{-\pi i
} disc_{s_{jl}}{\cal A}_{AB}^{A'B'+n},
\label{discontinuities}\ee $s_{jl}=(k_j+k_l)^2$  and in
the R.H.S. of (\ref{deriv}) the amplitude is considered
as a function of $y_i, \;\; i=0, \dots, n+1$, and
transverse momenta. Taking sum of the equations over $j$
from $0$ to $n+1$ it is easy to see from (\ref{deriv})
that  $\Re {\cal A}_{AB}^{A'B'+n} (y_{i})$ depends only
on differences of the rapidities $y_i$, as it must be.

The important point is that the relations (\ref{deriv}),
(\ref{discontinuities}) give a possibility {to find in
the NLA real parts of all MRK amplitudes in all orders of
coupling constant}, if $\Re {\cal A}_{AB}^{A'B'+n}
(y_{i})$ are known (for all $n$) in the one-loop
approximation. Indeed, these relations express all
partial derivatives of the real parts at some number $L$
of loops {through the discontinuities}, which can be
calculated using the $s$-channel unitarity in terms of
amplitudes with smaller number of loops; at that in the
NLA only the MRK is important and only real parts of
amplitudes do contribute. To find $\Re {\cal
A}_{AB}^{A'B'+n} (y_{i})$ besides the derivatives
determined by (\ref{deriv}), (\ref{discontinuities})
initial conditions are required; but since they can but
taken at fixed $y_i$, they are necessary only in the
one-loop approximation. Thus (\ref{deriv}),
(\ref{discontinuities}) allows to calculate $\Re {\cal
A}_{AB}^{A'B'+n} (y_{i})$ loop--by--loop using the
one-loop approximation as an input. Note that requirement
of equality of mixed derivatives taken  in different
orders imposes strong restrictions on the input. If it is
self-consistent, it determines $\Re {\cal
A}_{AB}^{A'B'+n} (y_{i})$ unambiguously.

Therefore in order to prove the Reggeization in the NLA
it is sufficient to know that (\ref{A 2-2+n}) is valid in
the one-loop approximation and satisfies (\ref{deriv}),
(\ref{discontinuities}),  where the discontinuities
(\ref{discontinuities}) are calculated using (\ref{A
2-2+n}) in the unitarity relations.

\section{Bootstrap relations for the Reggeized
amplitudes}

Substituting (\ref{A 2-2+n}) in (\ref{deriv}), we obtain
the relations
\[
\sum_{l=j+1}^{n+1}\Delta_{{jl}}
-\sum_{l=0}^{j-1}\Delta_{{lj}} \,
 =\,\left(\omega(t_{j+1})-\omega(t_{j})\right)\Re
\;  {\cal A}_{AB}^{A'B'+n}~,\label{bootstrap relations}
\]
which are called {bootstrap relations}. Evidently, there
is an {infinite number of the bootstrap relations},
because there is an infinite number of  the amplitudes
${\cal A}_{AB}^{A'B'+n}$. At the first sight, {it seems a
miracle to satisfy all of them}, since all these
amplitudes are expressed through several Reggeon vertices
and the gluon Regge trajectory. Moreover, {it is quite
nontrivial to satisfy even some definite bootstrap
relation for a definite amplitude}, because it connects
two infinite series in powers of $y_i$, and therefore it
leads to an infinite number of equalities  between
coefficients of these series.

In fact, two miracles must occur in order to satisfy all
the bootstrap relations: first, {for each particular
amplitude ${\cal A}_{AB}^{A'B'+n}$ it must be possible to
reduce the bootstrap relation to a limited number of
restrictions} ({bootstrap conditions}) {on the gluon
trajectory and the Reggeon vertices}, and secondly,
starting from some $n=n_0$ {these bootstrap conditions
must be the same} as obtained for amplitudes with
$n<n_0$. And finally, {all bootstrap conditions must be
satisfied by known expressions for the trajectory and the
vertices}.

\section{Representation of the discontinuities}

The miracles start from  the discontinuities {calculated
using (\ref{A 2-2+n})} and the $s$--channel unitarity. To
present them in a compact form let us use operator
denotations. Then the values of
$(2\pi)^{D-1}\delta(q_{i\bot}-q_{(j+1)\bot}-\sum_{l=i}^{l=j}
k_{l\bot})\frac{\Delta_{{ij}}}{4N_c}$ can be presented
\cite{FKR-tbp} as obtained from the R.H.S. of (\ref{A
2-2+n}) by the replacement of
\begin{equation}
\gamma^{J_i}(q_i,q_{i+1})\left( \prod_{l=i+1}^j
 \frac{e^{\omega(q_i)(y_{l-1}-y_l)}}{q^2_l}\gamma^{J_l}(q_l,q_{l+1})
 \right) \label{sub1}
\end{equation}
for
\begin{equation}
\langle J_i, \bar R_i|\left( \prod_{l=i+1}^{j-1}
 e^{\hat{\cal K}(y_{l-1}-y_l)}\hat{\cal J}_l \right)
 e^{\hat{\cal K}(y_{j-1}-y_j)} |\bar J_j, \bar R_{j+1}\rangle.
 \label{sub2}
\end{equation}
Eqs. (\ref{sub1}), (\ref{sub2}) remain valid for $i=0$
with the substitutions
$\gamma^{J_0}(q_0,q_{1})\rightarrow \Gamma_{A^{\prime}
A}$ and $\langle J_0, \bar R_0|\rightarrow \langle
A^{\prime}, \bar A |$,  as well for  $j=n+1$,  with the
substitutions
$\gamma^{J_{n+1}}(q_{n+1},q_{n+2})\rightarrow
\Gamma_{B^{\prime} B}$ and $ |\bar J_{n+1}, \bar
R_{n+2}\rangle\rightarrow |\bar B^{\prime}, B \rangle$.
Here $\hat{\cal K}$ is the operator of colour octet BFKL
kernel, $\hat{\cal J}_i$ is the jet $J_i$ production
operator; the state $|\bar B^{\prime}, B \rangle$
represents the scattering particles (jets) $B$ and
${B}^{\prime}$ from the cross-channel point of view,
$|\bar J_{l}, \bar R_{l+1}\rangle $ represents  Reggeon
with momentum $q_{l+1}$ and  jet $J_{l}$; $\langle
{A}^{\prime}, \bar A|$ and  and $\langle  J_{l}, \bar
R_{l}|$ are corresponding conjugate states.  The matrix
elements in (\ref{sub2}) is calculated using the full set
of states $|r_{1\bot},r_{2\bot},a\rangle $ of two
Reggeons with definite transverse momenta in the adjoint
representation.  If $|r_\bot,a\rangle$ is the
one--Reggeon state with transverse momentum $r_\bot$ and
colour index $a$, then $|r_{1\bot},r_{2\bot},a\rangle
=if_{aa_1a_2}|r_{1\bot},a_1\rangle
|r_{2\bot},a_2\rangle$. We use normalization $\langle
r_{1\bot},a_1|r_{2\bot},a_2\rangle=
\delta_{a_1a_2}r^2_{1\bot} \delta(r_{1\bot}-r_{2\bot})$.
Everywhere in the following symmetrization in Reggeon
momenta $r_1$ and $r_2$ is assumed.

At that we have \be\hat{\cal K}=\omega(\hat
r_1)+\omega(\hat r_2)+\hat{\cal K}_r, \label{kernel}\ee
where the subscript $r$ means the contribution coming
from real particle production. To escape a double
counting in the NLA we introduce an auxiliary parameter
$\Delta\gg 1$ dependence from which vanishes at large
$\Delta$:
 \be \hat{\cal K}_r=\hat{\cal
K}^{\Delta}_r-\hat{\cal K}^{B}_r\hat{\cal K}^{B}_r
\Delta; \ee  the superscript $B$ here and below denotes
quantities calculated in the LO; $\hat{\cal
K}^{\Delta}_r$ concerns with production of jets $J$ with
intervals of particle rapidities $\Delta_J$ in them less
than $\Delta$:
\[
\langle r'_{1\bot},r'_{2\bot},a'|\hat{\cal
K}^{\Delta}_r|r_{1\bot},r_{2\bot},a\rangle
=\delta_{aa'}\delta(r_{1\bot}+r_{2\bot}-r'_{1\bot}-r'_{2\bot})
\]
\begin{equation}
\times
\frac{f_{c_1c_2c}f_{c'_1c'_2c}}{N_c(N_c^2-1)}\sum_J\int
\gamma^J_{c_1c'_1}(r_1,r'_1)\left(\gamma^J_{c_2c'_2}(-r_2,-r'_2)
\right)^*\frac{d\phi_J}{2(2\pi)^{D-1}}\theta(\Delta-\Delta_J),
\label{real kernel}
\end{equation}
where
\begin{equation}
d\phi_{J}=\frac{d k^2_J}{2\pi} (2\pi)^{D}
\delta^D(k_J-\sum_i l_i)
\prod_{i}\frac{d^{D-1}l_{i}}{\left( 2\pi
\right)^{D-1}2\epsilon _{i}}~ \label{phi jet}
\end{equation}
for a jet $J$ with total momentum $k_J$ consisting of
particles with momenta $l_i$; $\;\;{\cal K}^{B}_r$ is
given by (\ref{real kernel}) in the LO, and the second
term in (\ref{kernel}) serves for subtraction of
contributions already taken into account in the LLA.

The states  describing particle (jet) transitions due to
interaction with Reggeized gluons are presented as
\begin{equation}
|\bar B^\prime B\rangle=|\bar B^\prime
B^{\Delta}\rangle-\left({\omega^B(\hat
r_1)}\ln\left|\frac{\hat r_{1\bot}}{q_{B\bot}}\right|
+{\omega^B(\hat r_2)}\ln\left|\frac{\hat
r_{2\bot}}{q_{B\bot}}\right|
 + \hat{\cal K}_r^B\;\Delta\right)|\bar B^\prime
B^{B}\rangle,  \label{BB}
\end{equation}
\begin{equation}
\langle r_{1\bot},r_{2\bot};a|\bar B^\prime
B^{\Delta}\rangle =
\delta(q_{B\bot}-r_{1\bot}-r_{2\bot})i\frac{f_{ac_1c_2}}{N_c}
\sum_{ \tilde B  }\int \Gamma^{c_1}_{\tilde B B}
\Gamma^{c_2}_{B' \tilde B }
 d\phi_{ \tilde B}\prod_l\theta(\Delta -(z_l-y_B))~,\label{BBdelta}
\end{equation}
and
\begin{equation}
\langle A^\prime \bar A|=\langle A^\prime \bar
A^\Delta|-\langle A^\prime \bar A^B|\left({\omega^B(\hat
r_1)}\ln\left|\frac{\hat r_{1\bot}}{q_{A\bot}}\right|
+{\omega^B(\hat r_2)}\ln\left|\frac{\hat
r_{2\bot}}{q_{A\bot}}\right|
 + \hat{\cal K}_r^B\;\Delta\right),  \label{AA}
\end{equation}
\begin{equation}
\langle A^\prime \bar
A^{\Delta}|r_{1\bot},r_{2\bot};a\rangle =
\delta(q_{A\bot}+r_{1\bot}+r_{2\bot})i\frac{f_{ac_1c_2}}{N_c}
\sum_{ \tilde A }\int \Gamma^{c_1}_{\tilde A A }
\Gamma^{c_2}_{A' \tilde A  }
 d\phi_{ \tilde A}\prod_l\theta(\Delta -(y_A-z_l))~,\label{AAdelta}
\end{equation}
where $q_{A}=p_A-p_{A^\prime}$, $q_{B}=p_B-p_{B^\prime}$
and  $z_l$ are rapidities of particles in intermediate
jets. Note that when a two-particle jet enters in some
state, the second term in corresponding equation  can be
omitted and the first taken in the Born approximation.

Quite analogously
\[
 |\bar J_i,\bar R_{i+1}\rangle=|\bar J_i,\bar R_{i+1}^\Delta\rangle
 -\left(\left({\omega(\hat
r_1)}-{\omega(q_{i+1})}\right)\ln\left|\frac{k_{i\bot}}
{(\hat r_{1\bot}+q_{(i+1)\bot})}\right| \right.
\]
\begin{equation}
\left.
+{\omega(\hat r_2)}\ln\left|\frac{k_{i\bot}}{\hat
r_{2\bot}}\right|
 + \hat{\cal K}_r^{Born}\Delta\right)|\bar J_i,\bar R_{i+1}^B\rangle.
\end{equation}
\[
\langle r_{1\bot},r_{2\bot};a|\bar J_i,\bar
R_{i+1}^\Delta\rangle =
\delta(q_{(i+1)\bot}+k_{i\bot}+r_{1\bot}+r_{2\bot})
\]
\begin{equation}
\times i\frac{f_{ac_1c_2}}{N_c}\sum_{ J}\int \gamma^{J
}_{c_1a_{i+1}}(-r_1, q_{i+1}) \Gamma^{c_2}_{J_i J }
 d\phi_{ J }\prod\theta(\Delta-(z_l-y_i),
\end{equation}
and
\[
\langle J_i, \bar R_i|=\langle J_i, \bar
R_i^\Delta|-\langle J_i, \bar R_i^
B|\left(\left({\omega(q_i)}-\omega(\hat
r_1)\right)\ln\left|\frac{k_{i\bot}} {(q_{i\bot}+\hat
r_{1\bot})}\right|\right.
\]
\begin{equation}
\left.-{\omega(\hat r_2)}\ln\left|\frac{k_{i\bot}}{\hat
r_{2\bot}}\right|
 + \hat{\cal K}_r^{Born}\Delta\right),
\end{equation}
\[
\langle J_i \bar
R_i^{\Delta}|r_{1\bot},r_{2\bot};a\rangle =
\delta(r_{1\bot}+r_{2\bot}+q_{i\bot}-k_{i\bot})
i\frac{f_{ac_1c_2}}{N_c}
\]
\begin{equation}
\times\sum_{ J}\int \gamma^{J }_{a_ic_1}(q_i, -r_1)
\Gamma^{c_2}_{J_i J }
 d\phi_{ J }\prod\theta(\Delta -(y_i-z_l)),
\end{equation}
where $a_{l}$ are colour indices  of Reggeons $R_l$.

At last,
\[
\hat{\cal J}_i=\hat{\cal J}_i^{\Delta}-\left( \hat{\cal
K}\hat{\cal J}_i+\hat{\cal J}_i\hat{\cal
K}\right)\;\Delta, \;\;\; \langle
r^\prime_{1\perp},r^\prime_{2\perp}; a^\prime|\hat{\cal
\cal J}_i|r_{1\perp},r_{2\perp}; a\rangle
\]
\[=\delta(r_{1\perp}+r_{2\perp}-k_{i\perp}-r^\prime_{1\perp}-
r^\prime_{2\perp}) \frac{f^{ac_1c_2}f^{a^\prime
c^\prime_1c^\prime_2}}{N_c} \left[2
\gamma^{J_i}_{c_1c^\prime_1}(r_{1},
r^\prime_1)\delta(r_{2\perp}-r^\prime_{2\perp})
 r_{2\perp}^{~2}\delta_{c_2c_2^\prime}\right.
\]
\begin{equation}
\left.+\sum_{J }\int_{y_i-\Delta}^{y_i+\Delta}
\frac{dz_{J}}{2(2\pi)^{D-1}}\left(
 \gamma^{J}_{c_1c^\prime_1}(r_1, r^\prime_1)
 \left(\gamma_{c_2c_2\prime}^{\bar J_i J}
(- r_{2},- r^\prime_{2})\right)^*+\gamma^{ J_i
J}_{c_1c^\prime_1}(r_1, r^\prime_1)
 \left(\gamma_{c_2c_2\prime}^{J}
(- r_{2},- r^\prime_{2})\right)^*\right)\right].
\label{jet-delta}
\end{equation}
When $J_i$ is a two-particle jet, in the NLA in the first
of these equations  the second term  can be omitted and
the first taken in the Born approximation, so that the
last term in (\ref{jet-delta}) must be retained only when
both $J_i$ and $J$ are  single gluons.

All Reggeon vertices entering in these equations, as well
as the gluon trajectory,  are known now with required
accuracy (see \cite{VF03} and references therein).

\section{Bootstrap conditions}

Using representation (\ref{sub2}) for the discontinuities
it was proved \cite{FKR-tbp} that an infinite number of
the bootstrap relations (\ref{bootstrap relations}) are
satisfied if the following bootstrap conditions are
fulfilled: the impact factors for scattering particles
satisfy equations \be |\bar B' B\rangle =
\frac{g}{2}\Gamma_{B'B}|R_{\omega}(q_B)\rangle, \;\;
\langle{{A'\bar A}}| =\frac{g}{2}\Gamma_{A'A}\langle
R_{\omega}(q_A)|, \label{second bootstrap}\ee where
$|R_{\omega}(q)\rangle $ is the universal (process
independent) eigenstate of the kernel $\hat {\cal K}$
with the eigenvalue $\omega(t)$,  \be
 \hat {\cal K}|R_{\omega}(q)\rangle = \omega(q)
 |R_{\omega}(q)\rangle, \label{first bootstrap}
\ee and the normalization \be \frac{g^2 t
N_c}{2(2\pi)^{D-1}}\langle
R_{\omega}(q)|R_{\omega}(q)\rangle=\omega(t)~; \ee  the
Reggeon-gluon impact factors and the gluon production
vertices satisfy the equations
\[
|\bar J_i \bar
R_{i+1}\rangle+\frac{gq_{i+1}^2}{2}\hat{\cal
J}_{i}|R_{\omega}(q_{i+1})\rangle  =
\frac{g}{2}\gamma^{J_{i}}(q_{i},q_{i+1})|R_{\omega}(q_{i})\rangle
, \;
\]
\be \langle J_i\bar R_i|+ \frac{gq_i^2}{2}\langle
R_{\omega}(q_i)|\hat{\cal J}_i =
\frac{g}{2}\gamma^{J_i}(q_i,q_{i+1})\langle
R_{\omega}(q_{i+1})|. \label{third bootstrap} \ee
Actually the second of equations (\ref{second
bootstrap}), (\ref{third bootstrap}) are not independent;
they follow from the first ones.

The bootstrap conditions  (\ref{second bootstrap}) and
(\ref{first bootstrap}) are known for a long time
\cite{FF98}$^-\;$\cite{FFKP00} and are proved to be
satisfied \cite{FFKPIF}$^-\;$\cite{FP02}.  The bootstrap
relations for {elastic amplitudes} require only a weak
form of the conditions (\ref{second bootstrap}) and
(\ref{first bootstrap}), namely these conditions
projected on $R_\omega$.   It was recognized
\cite{V.F.02} that the bootstrap relations for {one-gluon
production amplitudes}  besides (\ref{second bootstrap})
and (\ref{first bootstrap}) require also a weak form of
the condition (\ref{third bootstrap}). Thus, {the
bootstrap relations for one-gluon production amplitudes
play a twofold role: they strengthen the conditions
imposed by the elastic bootstrap and give a new one}. One
could expect that the history will repeat itself upon
addition of each next gluon in the final state. If it
were so, we would have to consider the bootstrap
relations for production of arbitrary number of gluons
and would obtain an infinite number of bootstrap
conditions. Fortunately, history is repeated only partly:
it occurs that already {the bootstrap relations for
two-gluon production only require the strong form of the
last condition (i.e. (\ref{third bootstrap})) and don't
require new conditions} \cite{BFF03}. At last, it was
proved \cite{FKR-tbp} that all bootstrap relations
(\ref{bootstrap relations}) are satisfied if the
conditions (\ref{second bootstrap})-(\ref{third
bootstrap}) are fulfilled.

The bootstrap conditions with  two-particle jets are
required  in the NLA  only with the Reggeon vertices
taken in the Born approximation. They were checked and
proved to be satisfied in \cite{FKR03}$^,\;$ \cite{VF03}.
After that only (\ref{third bootstrap}) remained
unproved. Its fulfilment was proved recently
\cite{FKR-tbp}. Thus, now it is shown that all bootstrap
conditions are fulfilled, that completes the proof of the
gluon Reggeization.

\section{Summary}
The gluon Reggeization is one of remarkable properties of
QCD. It is extremely important for description of high
energy processes. In particular, it appears as the basis
of the BFKL approach to summation of the terms
strengthened by powers of $\log(1/x)$. The hypothesis is
extremely powerful, since all scattering amplitudes are
expressed  in terms of the gluon trajectory and several
Reggeon vertices. Now the hypothesis is proved in the
NLA. The proof is based on the bootstrap relations. It is
shown that  an infinite number of these relations is
reduced to several bootstrap conditions on the gluon
trajectory and the Reggeon vertices. It is shown that
fulfilment of these conditions means a proof of the
Reggeization hypothesis.  All bootstrap conditions are
formulated explicitly and are proved to be fulfilled.


\newpage


\begin{thebibliography}{99}
\bibitem{BFKL}
V.S. Fadin, E.A. Kuraev and L.N. Lipatov, Phys. Lett. B
{\bf 60}, 50 (1975); E.A. Kuraev, L.N. Lipatov and V.S.
Fadin, Zh. Eksp. Teor. Fiz. {\bf 71}, 840 (1976) [Sov.
Phys. JETP {\bf 44}, 443 (1976)]; Zh. Eksp. Teor. Fiz.
{\bf 72}, 377 (1977) [Sov. Phys. JETP {\bf 45}, 199
(1977)]; Ya.Ya. Balitskii and L.N. Lipatov, Yad. Fiz.
{\bf 28}, 1597 (1978) [Sov. J. Nucl. Phys. {\bf 28}, 822
(1978)].
\bibitem{BLF}
Ya.Ya.~Balitskii, L.N.~Lipatov and V.S.~Fadin, in {\it
Materials of IV Winter School of LNPI} (Leningrad, 1979)
p.109.
\bibitem{Kucs}
T. Kucs, arXiv:hep-ph/0403023;
\bibitem{V.F.02}
V.S.~Fadin,
{\it Talk given at the NATO Advanced Research Workshop
"Diffraction 2002",  August 31 - September 6, 2002,
Alushta, Crimea, Ukraine},  in {\it Diffraction 2002},
Ed. by R. Fiore {\it et al.}, NATO Science Series, Vol.
101, p.235.
\bibitem{FKR-tbp} V.S.~Fadin, M.G.~Kozlov and A.V.~Reznichenko,
to be published.
\bibitem{VF03}
V.~S.~Fadin,
Phys.\ Atom.\ Nucl.\  {\bf 66} 2017  (2003).
\bibitem{FF98}
V.S.~Fadin and R.~Fiore, Phys. Lett. B {\bf 440}, 359
(1998).
\bibitem{Braun99}
M.~Braun and G.P.~Vacca, Phys. Lett. B {\bf 454}, 319;
M.~Braun, hep-ph/9901447 (1999).
\bibitem{FFKP00}
V.S.~Fadin, R.~Fiore, M.I.~Kotsky and A.~Papa,
Phys. Lett. B {\bf 495}, 329 (2000).
\bibitem{FFKPIF}
V.S.~Fadin, R.~Fiore, M.I.~Kotsky and A.~Papa,
Phys.\ Rev.\ D {\bf 61}, 094005 (2000);
{\bf 61}, 094006 (2000).
\bibitem{BV00}
M.~Braun and G.P.~Vacca, Phys. Lett. B {\bf 477}, 156
(2000).
\bibitem{FFP99}
V.S.~Fadin, R.~Fiore and A.~Papa,
Phys. Rev. D {\bf 60}, 074025 (1999);  V.S.~Fadin,
R.~Fiore and M.I.~Kotsky,
Phys. Lett. B {\bf 494}, 100 (2000).
\bibitem{FP02}
V.~S.~Fadin and A.~Papa,
Nucl.\ Phys.\ B {\bf 640}, 309 (2002).
\bibitem{BFF03}
J.~Bartels, V.~S.~Fadin and R.~Fiore,
Nucl.\ Phys.\ B {\bf 672} 329 (2003).
\bibitem{FKR03} V.S.~Fadin, M.G.~Kozlov and A.V.~Reznichenko,
Yad.\ Fiz.\ {\bf 67} (2004) 377[Phys.\ Atom.\ Nucl.\ {\bf
67}  359 (2004)].


\end{thebibliography}
\end{document}